\documentclass[a4paper,12pt]{article}

\usepackage{hyperref}
\usepackage{subfigure}
\usepackage{amssymb, graphics}
\usepackage{pslatex}
\usepackage{ae}
\usepackage{mathrsfs}
\usepackage{graphicx}
\usepackage{bbm}
\usepackage{latexsym}
\usepackage{dsfont}
\usepackage{amsfonts}
\usepackage{amsmath}
\usepackage[super]{nth}
\usepackage{cancel}

\usepackage{epsfig}
\usepackage{subfigure}
\usepackage{graphicx}
\usepackage{amsmath}
\usepackage{cite}
\usepackage{verbatim}

\newcommand{\mathsym}[1]{{}}

\newcommand{\qed}{\nobreak \ifvmode \relax \else \ifdim\lastskip<1.5em 
\hskip-\lastskip \hskip1.5em plus0em minus0.5em \fi \nobreak 
\vrule height0.75em width0.5em depth0.25em\fi}

\newcommand{\tr}{\mbox{tr}}

\def\app#1#2{  \mathrel{    \setbox0=\hbox{$#1\sim$}    \setbox2=\hbox{      \rlap{\hbox{$#1\propto$}}      \lower1.1\ht0\box0    }    \raise0.25\ht2\box2  }}

\begin{document}
\begin{titlepage}
\begin{center}

{\large \bf {Can One have Significant Deviations from Leptonic $3\times 3$ Unitarity in the 
Framework of Type I Seesaw Mechanism?}}

\vskip 1cm

Nuno Rosa Agostinho $^{a}$  \footnote{nunorosaagostinho@ub.edu}, 
G. C. Branco  $^{b}$ \footnote{gbranco@tecnico.ulisboa.pt}, 
Pedro M. F. Pereira $^{b}$ \footnote{pedromanuelpereira@tecnico.ulisboa.pt},
M. N. Rebelo $^{b}$ \footnote{rebelo@tecnico.ulisboa.pt},
and J. I. Silva-Marcos $^b$\footnote{juca@cftp.ist.utl.pt},

\vspace{1.0cm}

$^a$Departament de Fis\'\i ca Qu\` antica i Astrof\' \i sica and Institut 
de Ciencies del Cosmos, \\
Universitat de Barcelona, Diagonal 647, E-08028 Barcelona, Spain 

\bigskip

$^{b}$Centro de F\'isica Te\'orica de Part\'iculas -- CFTP and Departamento de F\' \i sica\\
Instituto Superior T\'ecnico -- IST, Universidade de Lisboa, Av. Rovisco Pais nr. 1, \\
P-1049-001 Lisboa, Portugal \\

\end{center}

\vskip 3cm

\begin{abstract}
We address the question of deviations from $3\times 3$ unitarity of the leptonic mixing matrix
showing that, contrary to conventional wisdom, one may have significant deviations from 
unitarity in the framework of type I seesaw mechanism. In order for this scenario 
to be feasible, at least one of the heavy neutrinos must have a mass at the TeV scale,
while the other two may have much larger masses. We present specific examples where 
deviations from $3\times 3$ unitarity are sufficiently small to conform to all the present 
stringent experimental bounds but are sufficiently large to have the potential 
for being detectable at the next round of experiments.
\end{abstract}

\end{titlepage}

\section{Introduction}
The discovery of neutrino oscillations and at least two non-vanishing neutrino 
masses, provides clear evidence for Physics Beyond the Standard Model (SM). 
The simplest extension of the SM accommodating two non-vanishing neutrino masses 
involves the addition of at least two right-handed neutrinos. The most general 
gauge invariant Lagrangian includes a right-handed bare Majorana mass matrix $M$.
As a result, the scale of $M$ can be much larger than the 
electroweak scale, which leads to an elegant explanation for the smallness of 
neutrino masses, through the seesaw mechanism \cite{Minkowski:1977sc},
\cite{Yanagida:1979as}, \cite{Glashow}, \cite{GellMann:1980vs}, \cite{Mohapatra:1979ia}.
The seesaw mechanism necessarily implies violations from $3 \times 3$ unitarity of the 
Pontecorvo-Maki-Nakagawa-Sakata  (PMNS) matrix, as well as Z-mediated lepton 
flavour violating couplings. The introduction of these heavy right-handed neutrinos 
can also have profound cosmological implications since they are a crucial component 
of the Leptogenesis mechanism to create the observed Baryon Asymmetry of the Universe 
(BAU) \cite{Fukugita:1986hr}. Leptogenesis is a very appealing scenario \cite{reviews1} but for
heavy neutrinos with masses many orders of magnitude higher than the electroweak scale 
it is difficult or impossible to test it at low energies 
\cite{delAguila:2008hw}, \cite{Deppisch:2015qwa}. 
Furthermore, without a flavour model, CP violation at high energies, relevant for
Leptogenesis, cannot be related to CP violation at low energies \cite{Branco:2001pq},
\cite{Rebelo:2002wj}. This may be possible in the context of a flavour model, involving
symmetries which allow to establish a connection between high and low energies 
\cite{Branco:2002kt},\cite{Frampton:2002qc}, \cite{Branco:2002xf}, \cite{Branco:2005jr},         
\cite{Pascoli:2006ci}, \cite{Branco:2007nb}, \cite{Hagedorn:2016lva},
\cite{Fukugita:2016hzf}. \\

At this stage, it should be emphasised that within the seesaw type I framework, 
the observed pattern of neutrino masses and mixing does not require that all the 
heavy neutrino masses be much larger than the electroweak scale. In this paper, 
we carefully examine the question of whether it is possible, within the seesaw 
type I mechanism, to have  experimentally detectable violations of $3 \times 3$ 
unitarity, taking into account the present experimental constraints. In particular, 
we address the following questions:

i) In the seesaw type I mechanism, is it possible to have significant deviations from
$3\times 3$ unitarity of the leptonic mixing matrix? By significant, 
we mean deviations which are sufficiently small 
to conform to all present stringent experimental constraints on these deviations, 
but are sufficiently large to be detectable in the next round of experiments. 
These experimental constraints arise from bounds on rare processes.

ii) In the case the scenario described in (i) can indeed be realised within the framework 
of seesaw type I, what are the requirements on the pattern of heavy neutrino masses?

For definiteness, we will work in a framework where three right-handed neutrinos are added
to the spectrum of the SM.  Our analysis starts with the introduction of the unitary 
$6 \times 6$ mixing matrix ${\cal V}$, characterising all the leptonic mixing. We write this 
$6 \times 6$ mixing matrix in terms of four blocks of $3 \times 3$ matrices. Using unitarity 
of ${\cal V}$, we show that the full matrix ${\cal V}$ can be expressed in terms of only 
three blocks of $3 \times 3$ matrices. 
Then we apply these results to the diagonalisation of the $6 \times 6$ neutrino mass 
matrix, including both Dirac and Majorana mass terms.  Through the use of a 
specially convenient exact parametrisation of the $6 \times 6$ 
leptonic unitary mixing matrix, 
we evaluate deviations from $3 \times 3$ unitarity and derive the 
maximum value of the lightest heavy neutrino mass which is 
required in order to generate significant deviations from unitarity in the 
framework of the seesaw type I mechanism.

The paper is organised as follows. In the next section, we review the seesaw mechanism, 
define our notation 
and introduce a specially convenient exact parametrisation of the $6 \times 6$ 
leptonic mixing matrix ${\cal V}$. We evaluate the size of the deviations of 
$3 \times 3$ unitarity in the present framework and derive a constraint on the 
magnitude of the mass of the heavy Majorana neutrinos, in order to have 
significant deviations of unitarity. Numerical examples are given 
in section 3 and some of the derivations of our results are put in an Appendix. 
Finally we present our conclusions in the last section.

\section{Deviations from Unitarity in the Leptonic Sector}

\subsection{ Type I Seesaw mechanism}
In the context of the Type I seesaw mechanism, with only three right-handed neutrinos
added to the Lagrangian of the SM, the leptonic mass terms are given by:
\begin{eqnarray}
{\cal L}_m  &=& -[\overline{{\nu}_{L}^0} m \nu_{R}^0 +
\frac{1}{2} \nu_{R}^{0T} C \ M \nu_{R}^0+
\overline{l_L^0} m_l l_R^0] + h. c.  \nonumber \\
&=& - [\frac{1}{2}† n_{L}^{T} C \ {\cal M}^* n_L +
\overline{l_L^0} m_l l_R^0 ] + h. c.,
\label{lep}
\end{eqnarray}
There is no loss of generality in choosing a weak basis where $m_l$ is already
real and diagonal. The analysis that follows is performed in this basis.
The neutrino mass matrix $\cal M $ is a $6 \times 6$ matrix and has the form:
\begin{eqnarray}
{\cal M}= \left(\begin{array}{cc}  
 0  & m \\
m^T & M \end{array}\right) 
\label{calm}
\end{eqnarray}
This matrix is diagonalised by the unitary transformation 
\begin{eqnarray}
{\cal V}^T {\cal M}^* {\cal V} = {\cal D} \label{dgm}  \qquad 
\mbox{i.e.} \qquad {\cal V}^\dagger {\cal M} = {\cal D} {\cal V}^T 
\label{vmvd}
\end{eqnarray}
where
\begin{equation}
{\cal D}=\left(\begin{array}{cc}
d & 0 \\
0 & D \end{array}\right) 
\end{equation}
with $d = \mbox{diag}.(m_1, m_2, m_3)$ and $D = \mbox{diag}.(M_1, M_2, M_3)$ denoting
respectively, the light and the heavy Majorana neutrino masses. 
The unitary $6 \times 6$ matrix ${\cal V} $ is often denoted in the literature as:
\begin{equation}
{\cal V} =\left( 
\begin{array}{cc}
K & R \\ 
S & Z
\end{array}
\right)\  
\label{unit}
\end{equation}
where $K$, $R$, $S$ and $Z$ are $3\times 3$ matrices. 
For $K$ and $Z$ non singular, we may write
\begin{equation}
{\cal V} =\left( 
\begin{array}{cc}
K & 0 \\ 
0 & Z
\end{array}
\right) \left( 
\begin{array}{cc}
{1\>\!\!\!\mathrm{I}}\, & Y \\ 
-X & {1\>\!\!\!\mathrm{I}}\,
\end{array}
\right);\quad -X=Z^{-1}S;\quad Y=K^{-1}R  \label{u1}
\end{equation}
From the unitary relation ${\cal V} \ {\cal V} ^{\dagger}={1\>\!\!\!
\mathrm{I}}_{(6\times 6)}$, we promptly conclude that
\begin{equation}
Y=X^{\dagger }  \label{xy}
\end{equation}
The matrix ${\cal V} $ can thus be written:
\begin{eqnarray}
{\cal V} =\left(\begin{array}{cc}
K & KX^\dagger \\
-ZX & Z \end{array}\right) \label{eq15}
\end{eqnarray}
We have thus made clear that the unitary $6 \times 6$ matrix ${\cal V}$ 
can be expressed in terms of three 
independent $3 \times 3$ matrices. From the unitarity 
of ${\cal V} $, we obtain:
\begin{equation}
\begin{array}{c}
K\ \left( {1\>\!\!\!\mathrm{I}}\,+X^{\dagger }X\right) \ K^{\dagger }
={1\>\!\!\!\mathrm{I}} \\ 
\\ 
Z\ \left( {1\>\!\!\!\mathrm{I}}\,+X\ X^{\dagger }\right) \ Z^{\dagger }
={1\>\!\!\!\mathrm{I}}
\end{array}
\label{kt}
\end{equation}
showing that the matrix $X$ parametrizes the deviations from unitary
of the matrices $K$ and $Z$. More explicitly:
\begin{equation}
\begin{array}{c}
K\ K^{\dagger} = {1\>\!\!\!\mathrm{I}} - K\ X^{\dagger }X \ K^{\dagger}
\\
Z\ Z^{\dagger} = {1\>\!\!\!\mathrm{I}} - Z\ X X^{\dagger } \ Z^{\dagger}
\end{array}
\label{3lin}
\end{equation}

From Eq.~(\ref{dgm}) we derive:
\begin{eqnarray}
-X^\dagger Z^\dagger m^T = d K^T \label{13a} \\
K^\dagger m -X^\dagger Z^\dagger M = - d X^T Z^T  \label{13b} \\
Z^\dagger m^T = D X^* K^T  \label{13c} \\
XK^\dagger m + Z^\dagger M = D Z^T \label{13d} 
\end{eqnarray}
replacing $Z^\dagger m^T$ from Eq.~(\ref{13c}) into Eq.~(\ref{13a}) we get
\begin{eqnarray}
d = - X^T \ D \ X 
\label{dxdx}
\end{eqnarray}
which implies that:
\begin{eqnarray}
X = \pm i \sqrt{D^{-1}} O_c \sqrt{d}
\label{xc}
\end{eqnarray}
where $O_c$ is a complex orthogonal matrix, i.e., $O_c^T O_c = 1\>\!\!\!\mathrm{I}$, 
or explicitly:
\begin{eqnarray}
|X_{ij}| = \left| (O_c)_{ij} \sqrt{\frac{m_j}{M_i}} \right|
\end{eqnarray}
It should be stressed that the parametrisation of the $6 \times 6$ unitary matrix $\cal{V}$
given by  Eq.~(\ref{eq15}) has the especial property of allowing to connect in a 
straightforward and simple way the masses of the light and the heavy neutrinos through an 
orthogonal complex matrix $O_c$, as can be seen from Eqs.~(\ref{dxdx}) and (\ref{xc}).
This is an important new result which plays a crucial r\^ ole in our analysis.

Since  $O_c$ is an orthogonal complex matrix, not all of its elements need to be small;
furthermore, not all the $M_i$ need to be much larger than the electroweak scale, in order 
for the seesaw mechanism to lead to naturally suppressed neutrino masses. 
These observations about the size of the elements of X are specially relevant 
in view of the fact that some of the important physical implications of the seesaw model
depend crucially on $X$. In particular, the deviations from $3 \times 3$ unitarity are controlled 
by $X$, as shown in Eq.~(\ref{3lin}).

Given the importance of the matrix $X$, one may ask whether it is possible to write the 
$6 \times 6$ unitary matrix ${\cal V}$ in terms of $3 \times 3$ blocks, where only
$3 \times 3$ unitary matrices enter, together with the matrix $X$. In the Appendix, 
we show that this is indeed possible, and that the matrix ${\cal V}$ can be written:
\begin{equation}
{\cal V}=\left( 
\begin{array}{cc}
K & R \\ 
S & Z
\end{array}
\right) =\left( 
\begin{array}{cc}
\Omega \left( \sqrt{{1\>\!\!\!\mathrm{I}}\,+X^{\dagger }X}\ \right)^{-1} 
& \Omega \left( \sqrt{{1\>\!\!\!\mathrm{I}}
\,+X^{\dagger }X}\ \right)^{-1}X^{\dagger } \\ 
- \Sigma \left( \sqrt{{1\>\!\!\!\mathrm{I}}\,+X\ X^{\dagger }}\
\right)^{-1}X\  & \Sigma  \left( \sqrt{{1\>\!\!\!\mathrm{I}}
\,+X\ X^{\dagger }}\ \right)^{-1}
\end{array}
\right)  \label{v1}
\end{equation}
where $\Omega$ and $\Sigma$ are $3 \times 3$ unitary matrices given by:
\begin{equation}
\Omega = U_{K}U^{\dagger } \qquad \Sigma = W_{Z}\ W^{\dagger }
\end{equation}
and $U$, $W$ are the unitary matrices that diagonalise respectively 
$X^\dagger  X$ and $X  X^\dagger$:
\begin{equation}
U^\dagger \ X^{\dagger }X \ U = d_{X}^{2}; \qquad
W^\dagger X X^\dagger \ W = d_{X}^{2}
\label{uwxx}
\end{equation}
It is also shown in the Appendix, that $U_K$ and $W_Z$ defined by:
\begin{equation}
U_{K} \equiv K\ U\ \sqrt{\left( {1\>\!\!\!\mathrm{I}}
\,+d_{X}^{2}\right) } \qquad 
W_Z \equiv Z\ W\ \sqrt{\left( {1\>\!\!\!\mathrm{I}}
\,+d_{X}^{2}\right) }
\label{esta}
\end{equation}
are in fact unitary matrices.

As will be explained in the next section, this will allow us, in our analysis, 
to trade the matrix $K$ by the combination $U_{K}U^{\dagger}$ 
which we identify as the best fit for $U_{PMNS}$ derived under the assumption of 
unitarity, multiplied by the remaining factor that parametrises the deviations 
from unitarity.

\subsection{On the size of deviations from unitarity}
In the framework of the type I seesaw, it is the block $K$ of 
the matrix ${\cal V}$ 
that takes the r\^ ole  played by
$U_{PMNS}$ matrix at low energies in models with only Dirac-type neutrino masses. 
Clearly, in this framework, $K$ is no longer a unitary 
matrix. However, present neutrino experiments are putting stringent constraints
on the deviations from unitarity.
In our search for significant deviations from unitarity of $K$,
we must make appropriate choices for the matrix $X$ in order to comply with the
experimental bounds, while at the same time obtain deviations that are sizeable 
enough to be detected  experimentally in the near future. It is our aim to show that, 
contrary to common wisdom, we can achieve this result with at least one of the heavy 
neutrinos with a mass at the TeV scale, without requiring unnaturally small Yukawa 
couplings and still have light neutrino masses not exceeding one eV.

Deviations from unitarity \cite{Antusch:2006vwa}, \cite{FernandezMartinez:2007ms},\cite{Antusch:2014woa},
\cite{Fernandez-Martinez:2016lgt}, \cite{Blennow:2016jkn}
of $K$ have been parametrised as the product of an
Hermitian matrix by a unitary matrix \cite{Fernandez-Martinez:2016lgt}:
\begin{equation}
K = ({1\>\!\!\!\mathrm{I}} - \eta ) V
\label{khu}
\end{equation}
where $\eta $ is an Hermitian matrix with small entries. 
In order to identify the different components of our matrix $K$, given in Eq.~(\ref{v1}), 
with the parametrisation of Eq.~(\ref{khu}) we rewrite $K$ as:
\begin{equation}
K = U_{K}U^{\dagger }\left( \sqrt{{1\>\!\!\!\mathrm{I}}\,+X^{\dagger }X}\
\right) ^{-1} = \left[ U_{K}U^{\dagger }\left( \sqrt{{1\>\!\!\!\mathrm{I}}\,+X^{\dagger }X}\
\right) ^{-1} U U_{K}^{\dagger } \right]  U_{K}U^{\dagger } 
\label{kz}
\end{equation}
inside the square brackets  we wrote the Hermitian matrix that we identify with
$({1\>\!\!\!\mathrm{I}} - \eta )$, and which will parametrise the deviations from unitarity.
The matrix $V \equiv U_{K}U^{\dagger }$ is a unitarity matrix which is identified with $U_{PMNS}$ 
obtained from the standard parametrisation \cite{Patrignani:2016xqp} for a unitary matrix. 
One can also write:
\begin{equation}
\left[ U_{K}U^{\dagger }\left( \sqrt{{1\>\!\!\!\mathrm{I}}\,+X^{\dagger }X}\
\right) ^{-1} U U_{K}^{\dagger } \right] \equiv \left[ U_{K} \left(
\sqrt{{1\>\!\!\!\mathrm{I}}\,+ d^2_X }\
\right) ^{-1} U_{K}^{\dagger } \right] 
\label{eq24}
\end{equation}
where $d^2_X$ is a $3 \times 3$ diagonal matrix, introduced in Eq.~(\ref{uwxx}).
Identifying the second expression of Eq.~(\ref{eq24}) to $({1\>\!\!\!\mathrm{I}} - \eta )$
we derive:
\begin{equation}
\label{etaa}
\eta = {1\>\!\!\!\mathrm{I}} - U_{K} \left( \sqrt{{1\>\!\!\!\mathrm{I}}\,+ d^2_X }\
\right) ^{-1} U_{K}^{\dagger } \approx \frac{1}{2} U_K \ d^2_X \  U_{K}^{\dagger } 
\end{equation}
for small $d^2_X$ .
The matrix $U_{PMNS}$ is then fixed making use of the present best fit values obtained from 
a global analysis based on the assumption of unitarity. As pointed out in 
\cite{Fernandez-Martinez:2016lgt}, from the phenomenological point of
view it is very useful to parametrise $K$ with the unitary matrix on the right, due to the
fact that experimentally it is not possible to determine which physical light neutrino is 
produced, and therefore, one must sum over the neutrino indices. As a result, most observables 
depend on $K K^\dagger$ which depends on the following combination:
\begin{equation}
\left( K K^\dagger \right)_{\alpha \beta} = \delta_{\alpha \beta} - 
2 \eta_{\alpha \beta} + \mathcal{O}(\eta^2_{\alpha \beta})
\end{equation}
 
The standard parametrisation  for $U_{PMNS}$ is given by \cite{Patrignani:2016xqp}:
\begin{eqnarray}
U_{PMNS}=\left(
\begin{array}{ccc}
c_{12} c_{13} & s_{12} c_{13} & s_{13} e^{-i \delta}  \\
-s_{12} c_{23} - c_{12} s_{23} s_{13}   e^{i \delta}
& \quad c_{12} c_{23}  - s_{12} s_{23}  s_{13} e^{i \delta} \quad 
& s_{23} c_{13}  \\
s_{12} s_{23} - c_{12} c_{23} s_{13} e^{i \delta}
& -c_{12} s_{23} - s_{12} c_{23} s_{13} e^{i \delta}
& c_{23} c_{13} 
\end{array}\right) \cdot  P
\label{std}
\end{eqnarray}
with P given by
\begin{equation}
P=\mathrm{diag} \ (1,e^{i\alpha_{21}}, e^{i\alpha_{31}})
\end{equation}
where $c_{ij} = \cos \theta_{ij}$, $s_{ij} = \sin \theta_{ij}$ 
and $\delta$ is a Dirac-type 
CP violating phase, while $\alpha_{21}$, $\alpha_{31}$ denote Majorana 
phases. Neutrino oscillation experiments are not sensitive to these 
factorisable phases.

There are several groups performing global phenomenological fits on 
$\theta_{12}$, $\theta_{23}$, $\theta_{13}$ and $\delta$, as well as on
neutrino mass differences \cite{deSalas:2017kay}, 
\cite{Capozzi:2017ipn}, \cite{Esteban:2016qun}.

The specific bounds vary slightly from group to group.
For definiteness we present in Table 1 the present bounds on neutrino
masses and leptonic mixing from \cite{deSalas:2017kay}. The quantities
$\Delta m^2_{ij}$ are defined by $(m^2_i - m^2_j)$.
\begin{center}
\begin{table}[h]
\caption{Neutrino oscillation parameter summary from \cite{deSalas:2017kay}. 
For $\Delta m^2_{31}$, 
$\sin^2 \protect\theta_{23}$ , $\sin^2 \protect\theta_{13}$, and $\protect
\delta$ the upper (lower) row corresponds to normal (inverted) neutrino mass
hierarchy.
${}^a$There is a local minimum in the second octant, at  $\sin^2 \theta_{23}= 0.596 $
with $\Delta \chi^2 = 2.08$ with respect to the global minimum. 
${}^b$ There is a local minimum in the first octant, at  $\sin^2 \theta_{23}= 0.426 $
with $\Delta \chi^2 = 1.68$ with respect to the global minimum for IO.}
\label{reps}
\begin{tabular}{ccc}
\hline\hline
Parameter & Best fit & $1 \sigma $ range \\ \hline
$\Delta m^2_{21}$ $[10^{-5} eV^2 ] $ & 7.56 & 7.37 -- 7.75\\ 
$|\Delta m^2_{31}|$ $[10^{-3} eV^2 ] (NO)$ & 2.55 & 2.41 -- 2.59 \\ 
$|\Delta m^2_{31}|$ $[10^{-3} eV^2 ] (IO) $ & 2.49 & 2.45 -- 2.53 \\ 
$\sin^2 \theta_{12}$ & 0.321 & 0.305 -- 0.339 \\ 
$\sin^2 \theta_{23} (NO)$ & 0.430  & 0.412 -- $0.450^a$ \\ 
$\sin^2 \theta_{23} (IO)$ & 0.596 & 0.576 -- $0.614^b$ \\ 
$\sin^2 \theta_{13}$ (NO)& 0.02155 & 0.02080 --0.02245 \\ 
$\sin^2 \theta_{13}$ (IO) & 0.02140 & 0.02055 -- 0.02222 \\ 
$\delta$ (NO) & 1.40 $\pi $ & 1.20 --1.71 $\pi$ \\ 
$\delta$ (IO) & 1.44 $\pi$ & 1.70 --1.21 $\pi$ \\ \hline
\end{tabular}
\end{table}
\end{center}
In Ref.~\cite{Fernandez-Martinez:2016lgt} global constraints are derived on 
the matrix $\eta$ through a fit of twenty eight observables including
the W boson mass, the effective mixing weak angle $\theta_W$, several 
ratios of $Z$ fermionic decays, the invisible width of the $Z$, several
ratios of weak decays constraining EW universality, weak decays constraining 
CKM unitarity and some radiative lepton flavour violating (LFV) processes.
The final result is translated into:
\begin{eqnarray}
\left| 2 \eta_{\alpha \beta} \right| \leq \left(
\begin{array}{ccc}
2.5 \times 10^{-3} & 2.4 \times 10^{-5}  & 2.7 \times 10^{-3}    \\
2.4 \times 10^{-5} & 4.0 \times 10^{-4}  & 1.2 \times 10^{-3}   \\
2.7 \times 10^{-3}  & 1.2 \times 10^{-3}  & 5.6 \times 10^{-3}  
\end{array} \right)
\label{hen}
\end{eqnarray}
Ref.~\cite{Fernandez-Martinez:2016lgt} also compares these bounds with 
those of previous studies \cite{Antusch:2014woa}, \cite{Antusch:2015mia},
pointing out that in general there is good agreement. Variations in the 
scale of the masses of the heavy neutrinos lead to small effects and therefore,
do not significantly change our analysis.

\subsection{The elements of the neutrino Dirac mass matrix $m$ and
deviations from unitarity}

In this subsection, we show that there is a correlation among:
\begin{itemize}
\item The size of deviations from unitarity of the $3 \times 3$
leptonic mixing matrix.
\item The mass of the lightest heavy neutrino.
\end{itemize}
From Eq.~(\ref{13c}) we get 
\begin{eqnarray}
m= K X^\dagger D {Z^*}^{-1}
\label{mmtu}
\end{eqnarray}
the experimental fact that $K$ is almost unitary implies that $Z$ is also
almost unitary. Therefore the Dirac mass matrix 
$m$ is of the same order as $X$ times $D$. Notice 
that the scale of $D$ may be of the order of the top quark mass, so that indeed
the Yukawa couplings need not be extremely small.  

The elements of the neutrino Dirac mass 
matrix $m$ are connected to the deviations from unitarity of the $3 \times 3$
leptonic mixing matrix. 
From Eq.~(\ref{mmtu}) together with Eqs.~(\ref{esta}), (\ref{kz}) and     (\ref{eq24}), we obtain:
\begin{equation}
m = U_{K}\left( \sqrt{\left( {1\>\!\!\!\mathrm{I}}\,+d_{X}^{2}\right) }\ \right)
^{-1}d_{X}\ W^{\dagger }\ D\ W^{\ast }\left( \sqrt{\left( {1\>\!\!\!\mathrm{I
}}\,+d_{X}^{2}\right) }\ \right) W_{Z}^{T} \label{mss}
\end{equation}
where we have used $d_{X}=$ $W^{\dagger }\ X\ U$ from Eq. (\ref{uwxx}).
Thus, we find
\begin{equation}
Tr\left[ mm^{\dagger}\right] =
Tr\left[ \left( \sqrt{\left( {1\>\!\!\!\mathrm{I}}\,+d_{X}^{2}\right) }\
\right) ^{-1}d_{X}\ W^{\dagger }\ D\ W^{\ast }\ \left( {1\>\!\!\!\mathrm{I}}%
\,+d_{X}^{2}\right) \ W^{T}\ D\ W\ d_{X}\ \left( \sqrt{\left( {1\>\!\!\!
\mathrm{I}}\,+d_{X}^{2}\right) }\ \right) ^{-1}\right] 
\label{mmt}
\end{equation}
As previously emphasised, deviations from $3 \times 3$ unitarity in $U_{PMNS}$
are controlled by the matrix $X$, as it is clear from Eq.~(\ref{3lin}). For
$X=0$, there are no deviations from unitarity. Small deviations from unitarity
correspond to $d_X$ small and, in that case, one has to a very good 
approximation,
\begin{equation}
Tr\left[ mm^{\dagger}\right] =
Tr\left[ d_{X}\ W^{\dagger }D^{2}W\ d_{X}\right] 
\end{equation}
which can be written as:
\begin{equation}
\begin{array}{l}
Tr\left[ mm^{\dagger}\right] =
d_{X_{1}}^{2}\left( M_{1}^{2}\left\vert W_{11}\right\vert
^{2}+M_{2}^{2}\left\vert W_{21}\right\vert ^{2}+M_{3}^{2}\left\vert
W_{31}\right\vert ^{2}\right) + \\ 
d_{X_{2}}^{2}\left( M_{1}^{2}\left\vert W_{12}\right\vert
^{2}+M_{2}^{2}\left\vert W_{22}\right\vert ^{2}+M_{3}^{2}\left\vert
W_{32}\right\vert ^{2}\right) + \\ 
d_{X_{3}}^{2}\left( M_{1}^{2}\left\vert W_{13}\right\vert
^{2}+M_{2}^{2}\left\vert W_{23}\right\vert ^{2}+M_{3}^{2}\left\vert
W_{33}\right\vert ^{2}\right)
\end{array}
\label{mmt1}
\end{equation}
It can be shown that, using the properties of orthogonal complex matrices, only one of the 
$d_{X_i}$, corresponding to $d_{X_3}$, can have a significant value (e.g.
$d_{X_3} \approx 10^{-2}$),  while the other two are negligible.
Thus, we find in good approximation
\begin{equation}
Tr\left[ mm^{\dagger}\right] =
d_{X_{3}}^{2}\left( M_{1}^{2}\left\vert W_{13}\right\vert
^{2}+M_{2}^{2}\left\vert W_{23}\right\vert ^{2}+M_{3}^{2}\left\vert
W_{33}\right\vert ^{2}\right)  \label{mmt2}
\end{equation}
or using the unitary of $W$
\begin{equation}
Tr\left[ mm^{\dagger}\right] =
d_{X_{3}}^{2}M_{1}^{2}\left( 1+\left( \frac{M_{2}^{2}}{M_{1}^{2}}-1\right)
\left\vert W_{23}\right\vert ^{2}+\left( \frac{M_{3}^{2}}{M_{1}^{2}}
-1\right) \left\vert W_{33}\right\vert ^{2}\right) 
\label{mmt3}
\end{equation}
which, with the choice $M_{3}\geq M_{2}\geq M_{1}$, leads to
\begin{equation}
d_{X_{3}}^{2}M_{1}^{2}\leq Tr\left[ mm^{\dagger }\right] =\sum_{i,j} \left\vert
m_{ij}\right\vert ^{2}   \label{dim}
\end{equation}
From Eq.~(\ref{dim}), it is clear that for significant values of $d_{X_3}$,
$M_1$ cannot be too large in order to avoid a too large value of 
$Tr\left[ mm^{\dagger}\right] $, which in turn would imply that 
at least one of the $\left\vert m_{ij}\right\vert ^{2}$ is too large.
This can be seen in both Fig.~1 and Fig.~3 where we plot $\frac{1}{2}d_{X_{3}}^{2}$ versus $M_1$.
Significant values of $d_{X_{3}}^{2}$  can only be obtained for
$M_1 \leq 1 - 2$ TeV.
\begin{figure}[h!]

\includegraphics[width=1.\linewidth]{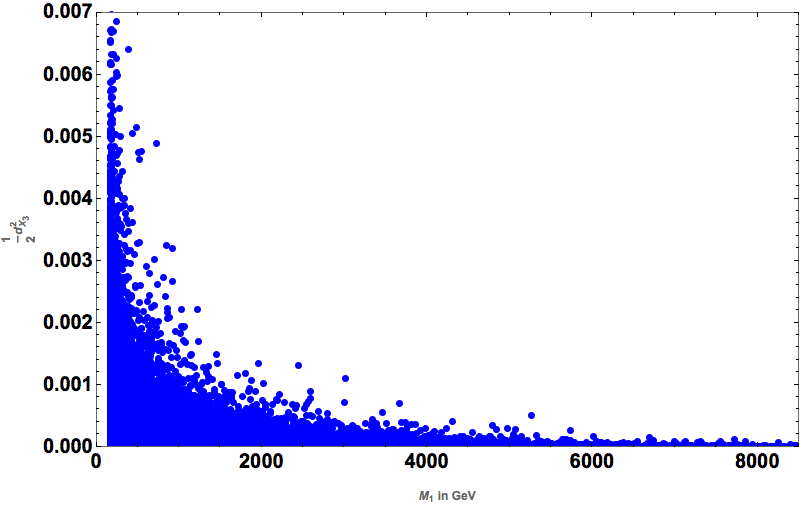}
  
\caption{Maximum deviations from unitarity as a function of $M_1$, generated under 
the condition that Tr$( mm^\dagger)\leq m_t^2$ and $|\eta_{12}|\leq 10^{-4}$.  }
\label{fig:sub1}

\end{figure}

In all our plots, we require $Tr\left[ mm^{\dagger }\right] \leq
m_{t}^{2}$. We consider the case of normal ordering 
and vary over the values of light neutrinos masses
$m_{i}$, up to $m_3 = 0.5$ eV.  Concerning  
the heavy Majorana masses $M_i$, we allow $M_3$ to reach values of the order of
$10^4 m_t$ and allow for all possible
forms of $O_{c}$. In Figs.~1,~2 we impose the condition that $\left\vert \eta _{12}\right\vert \leq 10^{-4}$, while in Figs.~3,~4, we chose
$\left\vert \eta _{12}\right\vert \leq 2\times 10^{-5}$.

In Figs.~2,~4, we also plot the $|\eta_{11}|$ deviations from unitarity.

\begin{figure}[h!]

\includegraphics[width=1.\linewidth]{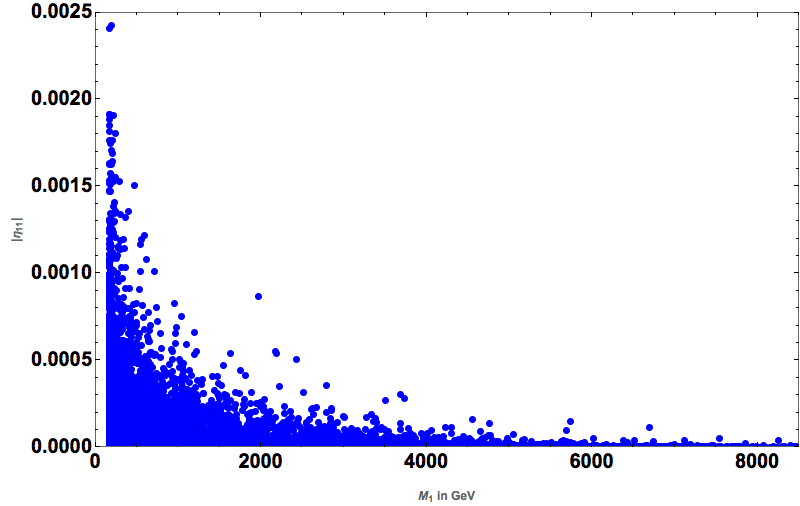}
  
\caption{$|\eta_{11}|$ deviations from unitarity as a function of $M_1$, generated under the condition that Tr$(mm^\dagger)\leq m_t^2$ and $|\eta_{12}|\leq 10^{-4}$. }
\label{fig:sub2}

\end{figure}

\begin{figure}[h!]

\includegraphics[width=1.\linewidth]{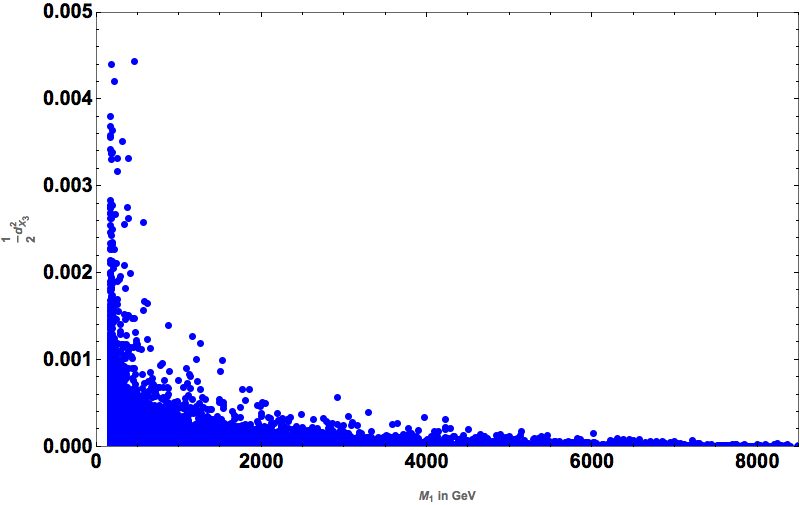}
  
\caption{Maximum deviations from unitarity as a function of $M_1$, generated under the condition 
that Tr$(mm^\dagger)\leq m_t^2$ and $|\eta_{12}|\leq 2\times 10^{-5}$.  }
\label{fig:sub3}

\end{figure}

\begin{figure}[h!]

\includegraphics[width=1.\linewidth]{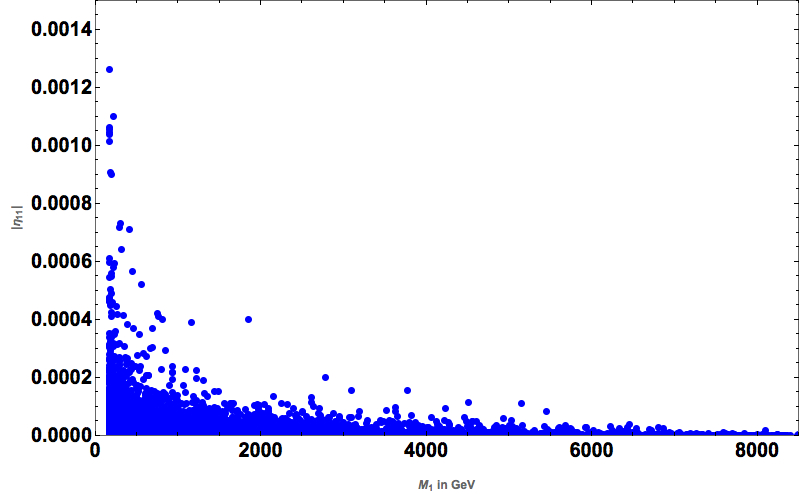}
  
\caption{$|\eta_{11}|$ deviations from unitarity as a function of $M_1$, generated under the condition that Tr$(mm^\dagger)\leq m_t^2$ and $|\eta_{12}|\leq 2\times 10^{-5}$. }
\label{fig:sub4}

\end{figure}

\section{Numerical Examples}

In this section, we present some illustrative results of our numerical analysis showing 
that it is possible to obtain the observed pattern of neutrino masses and mixing without 
requiring that all masses of the heavy Majorana neutrinos, $M_i$, be much larger 
than the electroweak scale.
We have realistic examples even when all the three heavy neutrinos have masses below 2 TeV. 
One might expect that lowering the scale of the heavy neutrino masses would 
result in the need for extremely small Yukawa couplings for the Dirac mass terms,  
thus defeating the rationale for the seesaw mechanism. However, this is not the case 
and, to illustrate, we include for each example the corresponding moduli of the entries 
of the neutrino Dirac mass matrices $m$ and the trace of the product $mm^\dagger$.  \\

\begin{table}[th]
\caption{Examples of deviations from unitarity, for the case of NH, expressed by the 
Hermitian matrix $\eta$.
We consider three different hierarchies for the heavy Majorana neutrino masses $M_i$. For 
each hierarchy we give two examples, where we vary the choice of the matrix $X$. The third 
column contains the Dirac-type neutrino mass matrix expressed in GeV units. The light 
neutrino  masses $m_i$ are given in eV.}
\begin{center}
\setlength{\tabcolsep}{0.5pc} 
\resizebox{\textwidth}{!}{\begin{tabular}{|c|c|c|c|c|c|}
\hline
Neutrino Masses \\ $m_1=0.005050$\ , \  $m_2=0.01005$\ , \  $m_3=0.05075$& $|\eta|$ &$|m|$\\
\hline
$\begin{pmatrix}
	M_1= 3 \ m_t\\
	M_2= 6\ m_t\\\
	M_3=9\ m_t\
\end{pmatrix}$ & 
$\begin{pmatrix}
	1.14\times 10^{-3} & 9.11\times 10^{-6}& 1.39\times 10^{-3} \\
	. & 7.27\times 10^{-8}& 1.11\times 10^{-5} \\
	.& . & 1.68\times 10^{-3}
\end{pmatrix}$
 & 
$\begin{pmatrix}
	35.183 &  6 \times 10^{-5} &  7 \times 10^{-11} \\
	0.280 & 2 \times 10^{-4} & 3 \times 10^{-10} \\
	42.707 & 1 \times 10^{-4} & 3 \times 10^{-10}
\end{pmatrix}$ \ \tr ($mm^\dagger$) = 0.102 $m^2_t$
\\
\hline

$\begin{pmatrix}
	M_1= 3 \ m_t\\
	M_2= 6\ m_t\\\
	M_3=9\ m_t\
\end{pmatrix}$ & 
$\begin{pmatrix}
	5.6\times 10^{-7} & 9.87\times 10^{-6}& 3.28\times 10^{-5} \\
	. &1.74\times 10^{-4}& 5.78\times 10^{-4} \\
	.& . & 1.92\times 10^{-3}
\end{pmatrix}$ &
$\begin{pmatrix}
	0.779 &  4 \times 10^{-5} & 1 \times 10^{-10}  \\
	13.735 & 2 \times 10^{-4} & 4 \times 10^{-10} \\
	45.63&  9 \times 10^{-5} & 3 \times 10^{-10}
\end{pmatrix}$ \ \tr ($mm^\dagger$) = 0.075 $m^2_t$

\\

\hline

$\begin{pmatrix}
	M_1=3\ m_t\ \\
	M_2=6\  m_t\\\
	M_3=1000\ m_t\
\end{pmatrix}$   &
$\begin{pmatrix}
	1.12\times 10^{-3} & 8.91\times 10^{-6}& 1.36\times 10^{-3} \\
	. & 7.10\times 10^{-8}& 1.08\times 10^{-5} \\
	.& . & 1.64\times 10^{-3}
\end{pmatrix}$
 & $\begin{pmatrix}
	34.779 &  2 \times 10^{-4} & 7 \times 10^{-11}  \\
	0.277 & 0.002 & 3 \times 10^{-10}  \\
	42.217& 0.002 & 3 \times 10^{-10} 
\end{pmatrix}$ \ \tr ($mm^\dagger$) =  0.099 $m^2_t$

\\

\hline

$\begin{pmatrix}
	M_1=3\ m_t\ \\
	M_2=6\  m_t\\\
	M_3=1000\ m_t\
\end{pmatrix}$   &
$\begin{pmatrix}
	5.6\times 10^{-7} & 9.87\times 10^{-6}& 3.28\times 10^{-5} \\
	. &1.74\times 10^{-4}& 5.78\times 10^{-4} \\
	.& . & 1.92\times 10^{-3}
\end{pmatrix}$
 & $\begin{pmatrix}
	0.779 &  5 \times 10^{-4}  & 1 \times 10^{-10} \\
	13.735 & 0.002 & 4 \times 10^{-10} \\
	45.643& 0.002 & 3 \times 10^{-10}
\end{pmatrix}$ \ \tr ($mm^\dagger$) = 0.075 $m^2_t$

\\

\hline

$\begin{pmatrix}
	M_1=3\ m_t\ \\
	M_2=60\ m_t\\\
	M_3=1000\ m_t\
\end{pmatrix}$   &
$\begin{pmatrix}
	1.09\times 10^{-3} & 8.71\times 10^{-6}& 1.33\times 10^{-3} \\
	. & 6.95\times 10^{-8}& 1.06\times 10^{-5} \\
	.& . & 1.61\times 10^{-3}
\end{pmatrix}$
 & $\begin{pmatrix}
	108.753 &   2 \times 10^{-4}  & 7 \times 10^{-11}  \\
	0.867 & 0.002 & 3 \times 10^{-10}  \\
	132.014 & 0.002 & 3 \times 10^{-10}  
\end{pmatrix}$ \ \tr ($mm^\dagger$) = 0.9708 $m^2_t$

\\
\hline

$\begin{pmatrix}
	M_1=3\ m_t\ \\
	M_2=60\ m_t\\\
	M_3=1000\ m_t\
\end{pmatrix}$   & 
$\begin{pmatrix}
	5.27\times 10^{-7} & 9.3\times 10^{-6}& 3.09\times 10^{-5} \\
	. &1.64\times 10^{-4}& 5.45\times 10^{-4} \\
	.& . & 1.81\times 10^{-3}
\end{pmatrix}$
 & $\begin{pmatrix}
	2.391 &  5 \times 10^{-4} & 1 \times 10^{-10}  \\
	42.159 & 0.002 & 4 \times 10^{-10} \\
	140.072& 0.002 & 3 \times 10^{-10} 
\end{pmatrix}$ \ \tr ($mm^\dagger$) = 0.710 $m^2_t$

\\

\hline
\end{tabular}}
\end{center}
\end{table}

In Table 2, we include examples with normal ordering of light neutrino masses, for 
a particular fixed value of these three masses, common to all examples. 
We consider three different hierarchies for the heavy Majorana neutrino masses $M_i$. For 
each choice of heavy neutrino mass hierarchy we give two examples, where we vary 
the matrix $O_c$. The other set of free parameters in our analysis are the 
Majorana-type phases entering in the choice of matrix $\Omega$.

In all examples given in Table 2 the orthogonal complex matrix is of the form:
\begin{equation}
O_{c}=\left( 
\begin{array}{ccc}
0 & \sqrt{x^{2}+1} & i\ x  \\ 
0 & i\ x & -\sqrt{x^{2}+1}  \\ 
1 & 0 & 0
\end{array}
\right) \  \left( 
\begin{array}{ccc}
\cos \tau & 0 & \sin \tau  \\ 
0 & 1 & 0  \\ 
-\sin \tau & 0 & \cos \tau
\end{array}
\right)
\label{OcO}
\end{equation}
with different choices of $x$ and $\tau$ for the different cases, respectively:
\begin{equation}
\begin{array}{cc}
x= 2.80 \times 10^{5}  & \ \tau = \frac{\pi}{2.9} \\
x= 2.48 \times 10^{5}  & \ \tau = \frac{\pi}{2.8} \\
x= 2.77 \times 10^{5}  & \ \tau = \frac{\pi}{2.9} \\
x= 2.48 \times 10^{5} & \ \tau = \frac{\pi}{2.8} \\
x= 3.27 \times 10^{5} & \ \tau = \frac{\pi}{2.9} \\
x= 2.88 \times 10^{5}  & \ \tau = \frac{\pi}{2.8} 
\end{array}
\end{equation}

The modulus of $m$ is obtained from Eq.~(\ref{mmtu}) and requires 
a choice for the $W_Z$ matrix. This choice has implications for the entries of the 
blocks $S$ and $Z$ of the matrix $\cal{V}$. Since, at the moment, there are no direct 
experimental constraints on these entries, providing guidance for this choice, we made
the simplest one, by fixing $W_Z$ to be equal to the identity. With a 
different choice of $W_Z$  we could in principle homogenise the orders of magnitude of
the entries of $m$ so that all of the Yukawa couplings would be of the same 
order of magnitude or close. \\

In Table 3 we include examples with inverted ordering of light neutrino masses, for 
a particular fixed value of these three masses, common to all examples.  We consider 
the same three different hierarchies for the heavy Majorana neutrino masses $M_i$, as
in the cases of Table 2. For each choice of heavy neutrino
mass hierarchy we give two examples, where we vary the matrix $O_c$. In the second,
fourth and sixth examples of Table 3, $O_c$ is of the form given by Eq.~(\ref{OcO}). 
In the first, third and fifth examples $O_c$ is of the form:
\begin{equation}
O_{c}=\left( 
\begin{array}{ccc}
\sqrt{x^{2}+1}& 0 & i\ x  \\ 
i\ x & 0 & -\sqrt{x^{2}+1}  \\ 
0 & 1 & 0
\end{array}
\right) \  \left( 
\begin{array}{ccc}
\cos \tau  & \sin \tau & 0  \\  
-\sin \tau &  \cos \tau & 0 \\
0 & 0 & 1
\end{array}
\right)
\label{1c1}
\end{equation}
The choices of $x$ and $\tau$ are respectively:
\begin{equation}
\begin{array}{cc}
x= 1.68 \times 10^{5} &  \ \tau = \frac{\pi}{10} \\
x= 9.65 \times 10^{4}  &  \ \tau = \frac{\pi}{3.9} \\
x= 1.68 \times 10^{5}  &  \ \tau = \frac{\pi}{10} \\
x= 9.65 \times 10^{4}  &  \ \tau = \frac{\pi}{3.9} \\
x=1.99 \times 10^{5}  &  \ \tau = \frac{\pi}{10} \\
x= 1.14 \times 10^{5}  & \ \tau = \frac{\pi}{3.9} 
\end{array}
\end{equation}
\begin{table}[th]
\caption{Examples of deviations from unitarity, for the case of IO, expressed by the 
Hermitian matrix $\eta$.
We consider three different hierarchies for the heavy Majorana neutrino masses $M_i$. For 
each hierarchy we give two examples, where we vary the choice of the matrix $X$. The third 
column contains the Dirac-type neutrino mass matrix expressed in GeV units. The light 
neutrino masses $m_i$ are given in eV.}
\begin{center}
\setlength{\tabcolsep}{0.5pc} 
\resizebox{\textwidth}{!}{\begin{tabular}{|c|c|c|c|c|c|}
\hline
Neutrino Masses \\ $m_1= 0.05064$\ , \  $m_2= 0.05138$\ , \  $m_3= 0.00864$& $|\eta|$ &$|m|$\\
\hline
$\begin{pmatrix}
	M_1= 3 \ m_t\\
	M_2= 6\ m_t\\\
	M_3=9\ m_t\
\end{pmatrix}$ & 
$\begin{pmatrix}
	1.17\times 10^{-3} & 9.96\times 10^{-6}& 1.2\times 10^{-3} \\
	. & 8.47\times 10^{-8}& 1.02\times 10^{-5} \\
	.& . & 1.22\times 10^{-3}
\end{pmatrix}$
 & 
$\begin{pmatrix}
	35.630 &  2 \times 10^{-4} &  5 \times 10^{-10} \\
	0.303 & 2 \times 10^{-4} & 2.5 \times 10^{-10} \\
	36.356 & 2 \times 10^{-4} & 3 \times 10^{-10}
\end{pmatrix}$ \ \tr ($mm^\dagger$) = 0.086 $m^2_t$
\\
\hline

$\begin{pmatrix}
	M_1= 3 \ m_t\\
	M_2= 6\ m_t\\\
	M_3=9\ m_t\
\end{pmatrix}$ & 
$\begin{pmatrix}
	1.32\times 10^{-7} & 5.34\times 10^{-6}& 1.08\times 10^{-5} \\
	. &2.15\times 10^{-4}& 4.35\times 10^{-4} \\
	.& . & 8.80\times 10^{-4}
\end{pmatrix}$ &
$\begin{pmatrix}
	0.379 &  2 \times 10^{-4} & 1 \times 10^{-9}  \\
	15.272 & 1 \times 10^{-4} & 3 \times 10^{-10} \\
	30.896 &  2 \times 10^{-5} & 2 \times 10^{-10}
\end{pmatrix}$ \ \tr ($mm^\dagger$) = 0.039 $m^2_t$

\\

\hline

$\begin{pmatrix}
	M_1=3\ m_t\ \\
	M_2=6\  m_t\\\
	M_3=1000\ m_t\
\end{pmatrix}$   &
$\begin{pmatrix}
	1.17\times 10^{-3} & 9.96\times 10^{-6}& 1.2\times 10^{-3} \\
	. & 8.47\times 10^{-8}& 1.02\times 10^{-5} \\
	.& . & 1.22\times 10^{-3}
\end{pmatrix}$

 & $\begin{pmatrix}
	35.630 &  2 \times 10^{-3} & 5 \times 10^{-10}  \\
	0.303 & 0.002 & 3 \times 10^{-10}  \\
	36.356 & 0.002 & 3 \times 10^{-10} 
\end{pmatrix}$ \ \tr ($mm^\dagger$) = 0.086 $m^2_t$

\\

\hline

$\begin{pmatrix}
	M_1=3\ m_t\ \\
	M_2=6\  m_t\\\
	M_3=1000\ m_t\
\end{pmatrix}$   &
$\begin{pmatrix}
	1.32\times 10^{-7} & 5.34\times 10^{-6}& 1.08\times 10^{-5} \\
	. &2.15\times 10^{-4}& 4.35\times 10^{-4} \\
	.& . & 8.80\times 10^{-4}
\end{pmatrix}$
 & $\begin{pmatrix}
	0.379 &  0.002  & 1 \times 10^{-9} \\
	15.272 & 0.001 & 3 \times 10^{-10} \\
	30.896 & 0.0004 & 2 \times 10^{-10}
\end{pmatrix}$ \ \tr ($mm^\dagger$) = 0.039 $m^2_t$

\\

\hline

$\begin{pmatrix}
	M_1=3\ m_t\ \\
	M_2=60\ m_t\\\
	M_3=1000\ m_t\
\end{pmatrix}$   &
$\begin{pmatrix}
	1.16\times 10^{-3} & 9.83\times 10^{-6}& 1.18\times 10^{-3} \\
	. & 8.36\times 10^{-8}& 1.0\times 10^{-5} \\
	.& . & 1.20\times 10^{-3}
\end{pmatrix}$
 & $\begin{pmatrix}
	111.946 &  0.002 & 5 \times 10^{-10}  \\
	0.952 & 0.002 & 3 \times 10^{-10}  \\
	114.227 & 0.002 & 3 \times 10^{-10}  
\end{pmatrix}$ \ \tr ($mm^\dagger$) = 0.849 $m^2_t$

\\
\hline

$\begin{pmatrix}
	M_1=3\ m_t\ \\
	M_2=60\ m_t\\\
	M_3=1000\ m_t\
\end{pmatrix}$   & 
$\begin{pmatrix}
	1.29\times 10^{-7} & 5.19\times 10^{-6}& 1.05\times 10^{-5} \\
	. &2.09\times 10^{-4}& 4.23\times 10^{-4} \\
	.& . & 8.56\times 10^{-4}
\end{pmatrix}$
 & $\begin{pmatrix}
	1.181 &  0.002 & 1 \times 10^{-9}  \\
	47.622 & 0.001 & 3 \times 10^{-10} \\
	96.343 & 0.0003 & 3 \times 10^{-10} 
\end{pmatrix}$ \ \tr ($mm^\dagger$) = 0.383 $m^2_t$

\\

\hline
\end{tabular}}
\end{center}
\end{table}

In all our examples the matrix $X$ 
will have several entries of order at most $10^{-2}$.

\section{Conclusions}
We have studied the possibility of having significant deviations from $3\times 3$ unitarity of
the leptonic mixing matrix in the framework of type-I seesaw mechanism.
The analysis was done in the framework of an extension of the Standard Model where 
three right-handed neutrinos are added to the spectrum. We have shown that the 
$6 \times 6$ 
unitary leptonic mixing matrix ${\cal V}$ can be written in terms of two unitary $3\times 3$ matrices 
and a matrix denoted X,  which controls the deviations from unitarity of the $3\times 3$ 
PMNS matrix. This parametrisation of the matrix ${\cal V}$,  played a crucial role in showing 
that one may have significant deviations from $3\times 3$ unitarity while conforming to all present 
data on neutrino masses and mixing, as well as respecting all stringent bounds on 
deviations from $3\times 3$ unitarity of the PMNS matrix. 

We have presented specific examples where the above deviations from unitarity are 
sufficiently large to have the potential for being observed at the next round 
of experiments. An important feature of our analysis is the fact that the mass of 
the lightest heavy sterile neutrino is, in principle, within experimental reach.
This can be achieved without unnaturally small neutrino Yukawa couplings, thus 
defying the conventional wisdom that heavy neutrino masses many orders of magnitude 
above the electroweak scale are needed for this purpose. 

\section*{Appendix}

We can write the Hermitian matrices $\left( {1\>\!\!\!\mathrm{I}}
\,+X^{\dagger }X\right) $ and $\left( {1\>\!\!\!\mathrm{I}}\,+X\ X^{\dagger
}\right) $ as, 
\begin{equation}
\begin{array}{c}
{1\>\!\!\!\mathrm{I}}\,+X^{\dagger }X=U\ \left( {1\>\!\!\!\mathrm{I}}
\,+d_{X}^{2}\right) \ U^{\dagger } \\ 
\\ 
{1\>\!\!\!\mathrm{I}}\,+X\ X^{\dagger }=W\ \left( {1\>\!\!\!\mathrm{I}}
\,+d_{X}^{2}\right) \ W^{\dagger }
\end{array}
\label{uw}
\end{equation}
where $U$ and $W$ are, by definition the unitary matrices
which diagonalise the Hermitian matrices $({1\>\!\!\!\mathrm{I}}\,+X^{\dagger}X)$
and $({1\>\!\!\!\mathrm{I}}\,+X\ X^{\dagger})$ respectively and 
$d_{X}^{2}$ is a diagonal matrix. Inserting Eq.~(\ref{uw}) into Eq.~(\ref{kt})
we obtain:
\begin{equation}
\begin{array}{c}
K\ U\ \left( {1\>\!\!\!\mathrm{I}}\,+d_{X}^{2}\right) \ U^{\dagger }\
K^{\dagger }=K\ U\ \sqrt{\left( {1\>\!\!\!\mathrm{I}}\,+d_{X}^{2}\right) }\
\cdot \sqrt{\left( {1\>\!\!\!\mathrm{I}}\,+d_{X}^{2}\right) }U^{\dagger }\
K^{\dagger }={1\>\!\!\!\mathrm{I}} \\ 
\\ 
Z\  W\ \left( {1\>\!\!\!\mathrm{I}}\,+d_{X}^{2}\right) \ W^{\dagger }\
Z^{\dagger }=Z\ W\ \sqrt{\left( {1\>\!\!\!\mathrm{I}}\,+d_{X}^{2}\right) }\
\cdot \sqrt{\left( {1\>\!\!\!\mathrm{I}}\,+d_{X}^{2}\right) }\ W^{\dagger }\
Z^{\dagger }={1\>\!\!\!\mathrm{I}}
\end{array}
\end{equation}
We therefore conclude that $K\ U\ \sqrt{\left( {1\>\!\!\!\mathrm{I}}
\,+d_{X}^{2}\right) }=U_{K}$ and $Z\ W\ \sqrt{\left( {1\>\!\!\!\mathrm{I}}
\,+d_{X}^{2}\right) }=W_{Z}$ are unitary matrices, and thus also that
\begin{equation}
\begin{array}{l}
K=U_{K}\ \left( \sqrt{\left( {1\>\!\!\!\mathrm{I}}\,+d_{X}^{2}\right) }\
\right) ^{-1}U^{\dagger }=U_{K}U^{\dagger }\left( U\left( \sqrt{\left( {%
1\>\!\!\!\mathrm{I}}\,+d_{X}^{2}\right) }\ \right) ^{-1}U^{\dagger }\right)
\\ 
\\ 
Z=W_{Z}\ \left( \sqrt{\left( {1\>\!\!\!\mathrm{I}}\,+d_{X}^{2}\right) }\
\right) ^{-1}W^{\dagger }=W_{Z}\ W^{\dagger }\left( W\left( \sqrt{\left( 
{1\>\!\!\!\mathrm{I}}\,+d_{X}^{2}\right) }\ \right) ^{-1}W^{\dagger }\right)
\end{array}
\label{kt1}
\end{equation}
or using the diagonalisation of the Hermitian matrices $\left( {1\>\!\!\!
\mathrm{I}}\,+X^{\dagger }X\right) $ and $\left( {1\>\!\!\!\mathrm{I}}\,+X\
X^{\dagger }\right) $ in Eq.(\ref{uw}), we write this, as
\begin{equation}
\begin{array}{c}
K=U_{K}\ \left( \sqrt{\left( {1\>\!\!\!\mathrm{I}}\,+d_{X}^{2}\right) }\
\right) ^{-1}U^{\dagger }=U_{K}U^{\dagger }\left( \sqrt{{1\>\!\!\!\mathrm{I}}
\,+X^{\dagger }X}\ \right) ^{-1} \\ 
\\ 
Z=W_{Z}\ \left( \sqrt{\left( {1\>\!\!\!\mathrm{I}}\,+d_{X}^{2}\right) }\
\right) ^{-1}W^{\dagger }=W_{Z}\ W^{\dagger }\left( \sqrt{{1\>\!\!\!\mathrm{I
}}\,+X\ X^{\dagger }}\ \right) ^{-1}\ 
\end{array}
\label{kt2}
\end{equation}
which lead to Eq.~(\ref{v1}).

\section*{Acknowledgments}
This work was partially supported by Funda\c c\~ ao para a Ci\^ encia e a 
Tecnologia (FCT, Portugal) through the projects CERN/FIS-NUC/0010/2015, 
CFTP-FCT Unit 777 \\ 
(UID/FIS/00777/2013) which are partially funded through 
POCTI (FEDER), COMPETE, QREN and EU. 
N.R.A has received funding/support from the European Union's 
Horizon 2020 research and innovation programme under the Marie 
Sklodowska-Curie grant agreement No 674896.
P.M.F.P. has a BIC-type Fellowship under project CERN/FIS-NUC/0010/2015.

\end{document}